\documentclass[]{spie}  

\usepackage{amsmath,amsfonts,amssymb}
\usepackage{graphicx}
\usepackage[colorlinks=true, allcolors=blue]{hyperref}

\usepackage{subfig}
\usepackage{hhline}
\usepackage{xkeyval,xcolor}
\makeatletter
\newlength{\sfp@hseplen}\newlength{\sfp@vseplen}
\define@cmdkey{subfigpos}[sfp@]{pos}[ul]{}
\define@cmdkey{subfigpos}[sfp@]{font}[\small]{}
\define@cmdkey{subfigpos}[sfp@]{vsep}[5pt]{\setlength{\sfp@vseplen}{\sfp@vsep}}
\define@cmdkey{subfigpos}[sfp@]{hsep}[5pt]{\setlength{\sfp@hseplen}{\sfp@hsep}}
\newcommand{\subfigimg}[3][,]{%
  \setkeys{Gin,subfigpos}{pos,font,vsep,hsep,#1}
  \setbox1=\hbox{\includegraphics{#3}}
  \ifnum\pdfstrcmp{\sfp@pos}{ul}=0
    \leavevmode\rlap{\usebox1}
    \rlap{\hspace*{\sfp@hsep}\raisebox{\dimexpr\ht1-\sfp@vsep}{\sfp@font{#2}}}
    \phantom{\usebox1}
  \else\ifnum\pdfstrcmp{\sfp@pos}{ur}=0
    \leavevmode\usebox1
    \llap{\raisebox{\dimexpr\ht1-\sfp@vsep}{\sfp@font{#2}}\hspace*{\sfp@hsep}}
  \else\ifnum\pdfstrcmp{\sfp@pos}{lr}=0
    \leavevmode\usebox1
    \llap{\raisebox{\sfp@vsep}{\sfp@font{#2}}\hspace*{\sfp@hsep}}
  \else
    \leavevmode\rlap{\usebox1}
    \rlap{\hspace*{\sfp@hseplen}\raisebox{\sfp@vsep}{\sfp@font{#2}}}
    \phantom{\usebox1}
  \fi\fi\fi
}
\makeatother

\title{A Graph-theoretic Algorithm for\\Small Bowel Path Tracking in CT Scans}

\author{Seung Yeon Shin}
\author{Sungwon Lee}
\author{Ronald M. Summers}
\affil{Imaging Biomarkers and Computer-Aided Diagnosis Laboratory\\Radiology and Imaging Sciences, National Institutes of Health Clinical Center, USA}

\authorinfo{Further author information: Send correspondence to Seung Yeon Shin (seungyeon.shin@nih.gov) or Ronald Summers (rms@nih.gov)}

\begin{document} 
\maketitle

\begin{abstract}

We present a novel graph-theoretic method for small bowel path tracking. It is formulated as finding the minimum cost path between given start and end nodes on a graph that is constructed based on the bowel wall detection. We observed that a trivial solution with many short-cuts is easily made even with the wall detection, where the tracked path penetrates indistinct walls around the contact between different parts of the small bowel. Thus, we propose to include must-pass nodes in finding the path to better cover the entire course of the small bowel. The proposed method does not entail training with ground-truth paths while the previous methods do. We acquired ground-truth paths that are all connected from start to end of the small bowel for 10 abdominal CT scans, which enables the evaluation of the path tracking for the entire course of the small bowel. The proposed method showed clear improvements in terms of several metrics compared to the baseline method. The maximum length of the path that is tracked without an error per scan, by the proposed method, is above $800mm$ on average.
 
\end{abstract}

\keywords{Small bowel path tracking, graph theory, shortest path, abdominal computed tomography.}

\section{INTRODUCTION}\label{sec:intro}

The small bowel is the longest (about 6 meters) part of the gastrointestinal tract between the stomach and the large bowel~\cite{gs20}. It is highly convoluted with its pliability so that it can fit into the abdominal cavity. As a result, the small bowel has many touching parts along its path, which makes it difficult to track from one end to another. Computed tomography (CT) has been regarded as a primary imaging modality for small bowel evaluation since it is safe and comfortable
compared to other imaging tests such as endoscopy~\cite{murphy14}. However, acquired 3D CT scans are reviewed by radiologists slice-by-slice for interpretation, which is laborious and time-consuming. An automatic system that identifies the small bowel could expedite the interpretation.

There have been research efforts to develop automatic methods for small bowel segmentation for the last decade~\cite{zhang13,oda20,shin20,shin21}. Considering the high difficulty of labeling the small bowel, the recent works based on deep learning presented data-efficient methods each by training with sparsely annotated CT volumes~\cite{oda20}, by incorporating a cylindrical shape prior~\cite{shin20}, or by developing an unsupervised domain adaptation technique for the small bowel~\cite{shin21}. While the segmentation is useful for detecting lesions, blockages, and for distinguishing the bowels from adjacent lesions in the mesentery, it may be insufficient to identify the whole structure of the small bowel especially for the parts with lumpy segmentation due to the aforementioned touching issue (Fig.~\ref{fig:qual_res} (A)).

Only a few previous works were performed to develop a method for small bowel path tracking~\cite{oda21,harten21}. In the work by Oda et al.~\cite{oda21}, the 3D U-Net~\cite{cicek16} is trained to predict a distance map from the centerlines of the small bowel. Qualitative evaluation was done for air-inflated bowels from CT scans, where the walls are more separable from the lumen than with other internal material. Harten et al.~\cite{harten21} developed a deep orientation classifier that predicts direction proposals to track the small bowel path in 3D cine-MR scans. The tracking was performed for parts of the small bowel within a given field-of-view of the scans rather than for the entire course of the small bowel. Both of the methods require ground-truth (GT) centerlines of the small bowel to train their neural network models.

In this paper, we present a novel graph-theoretic method for small bowel path tracking, which does not require any supervision on the centerlines. To prevent a computed path from incorrectly penetrating a bowel wall, we detect the walls by ridge detection based on which a graph is constructed. Given a start node and an end node, it is found in our experiment that simply finding the shortest path between them on the graph produces a trivial solution with short-cuts due to the imperfection of the wall detection (Fig.~\ref{fig:qual_res} (B)). We sample a set of must-pass nodes based on a predicted small bowel segmentation, and then find the optimal path that passes by all the must-pass nodes.

\section{METHOD}\label{sec:method}
\subsection{Dataset}

Our dataset consists of 30 high-resolution abdominal CT scans. All the scans are both intravenous and oral contrast-enhanced. Specifically, they were acquired with oral administration of Gastrografin and done during the portal venous phase. They were resampled to have isotropic voxels of $2mm^3$. Also, we manually cropped them along the z-axis to include from the diaphragm through the pelvis. Every scan has a GT segmentation, which is achieved using ``Segment Editor" module in 3DSlicer~\cite{fedorov12} by a radiologist with 12 years of experience. The GT segmentation covers the entire small bowel including the duodenum, jejunum, and ileum.

GT path of the small bowel is also achieved for a subset of 10 scans. It is drawn as interpolated curves that connect a series of manually placed points inside the small bowel. We note that this annotation took one full day for each scan. This set is used for the evaluation of the proposed path tracking method, and the remaining 20 scans, which we will call the segmentation training set, are used to train a network for segmentation prediction.

\subsection{Graph Construction}\label{subsec:graph_const}

\begin{figure}[t]
    \centering
    \includegraphics[width = 1\textwidth]{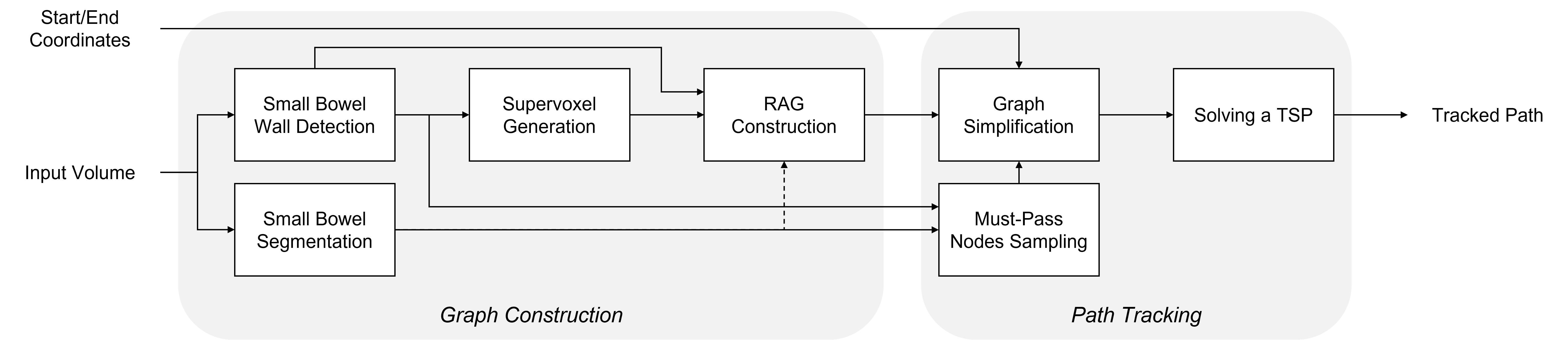}
    \caption{Block diagram of the proposed method for small bowel path tracking. A region adjacency graph (RAG) whose nodes and edges are built based on the wall detection is constructed. A set of must-pass nodes are sampled to better cover the entire course of the small bowel in finding the shortest path between given start and end nodes. Then, a simplified graph that consists of only the start and end nodes, and the sampled must-pass nodes is constructed. The path tracking is performed by solving a travelling salesman problem (TSP) on the simplified graph.}
    \label{fig:block_diagram} 
\end{figure}

The processing pipeline of the proposed method is summarized in Fig.~\ref{fig:block_diagram}.

\paragraph{Small Bowel Wall Detection and Supervoxel Generation}
It is desired that a computed path moves forward within the lumen while not penetrating bowel walls. In our dataset, the lumen appears brighter than the walls due to the administered oral contrast. To utilize this information, we detect the walls by ridge detection (valley detection to be exact). The Meijering filter~\cite{meijering04} is used in this paper.
Then, an input volume is segmented into supervoxels by Adaptive-SLIC~\cite{achanta12} to make a graph with the treatable number of nodes.
$S=\{s_i\}_{i=1}^{N}$ is the set of the generated supervoxels, where $N$ is the number of the supervoxels. The output of the wall detection is used as input for supervoxel generation so that the supervoxels would well adhere to boundaries between the lumen and the walls. 

\paragraph{RAG Construction}
A region adjacency graph (RAG), $G=(V,E)$, is constructed based on the generated supervoxels $S$ and the wall detection map. $V=\{v_i\}_{i=1}^{N}$ is a set of nodes and $E$ is a set of edges. Each node $v_i$ in the RAG corresponds to each supervoxel $s_i$. An edge $e_{i,j}$ is created between two adjacent supervoxels $s_i$, $s_j$ and the edge cost $C_{i,j}$ is defined as the average value in the wall detection map along their boundary. Thus, the cost between a supervoxel within the lumen but attached to the wall and a neighboring supervoxel on the wall would be high while that of adjacent supervoxels inside the lumen would be zero.

\paragraph{Small Bowel Segmentation}
In parallel with the wall detection, supervoxel generation, and the RAG construction, small bowel segmentation is also performed in this paper with two purposes: 1) to remove the graph nodes that are located outside the small bowel, and 2) to sample must-pass nodes within it. We will explain more about 2) in the following section. We adopt the baseline model used in Ref.~\citenum{shin20}, which is based on the 3D U-Net~\cite{cicek16}, for segmentation. Being trained using the segmentation training set, it is used to segment the small bowel for the set with GT paths.

\subsection{Path Tracking}

Given the graph $G$, we designate a start node $v_{st}$ and an end node $v_{ed}$ by providing coordinates for each on an input CT volume, possibly on the pylorus and the ileocecal junction, respectively.
Then, a simple approach that can be adopted for path tracking is to find the shortest path between the start and end nodes on $G$. From all the possible paths $\Gamma_G$ between them, the optimal path $\Gamma_G^* = \operatorname*{arg\,min}_{\Gamma_G} \sum_{e_{i,j} \in \Gamma_G} C_{i,j}$ is found by the Dijkstra’s algorithm.
However, we found in our experiment that it produces trivial solutions with many short-cuts over the walls due to the imperfection of the wall detection. 

\paragraph{Must-Pass Nodes Sampling}
We thus propose to use must-pass nodes in finding the shortest path to let the search better cover the entire course of the small bowel. Firstly, we compute the Euclidean distance transformation from the inverted small bowel segmentation and the wall detection, where each voxel value represents the smallest Euclidean distance to the walls. Then, the must-pass nodes are sampled as local peaks on the distance map with the hope that they would be as close as possible to the centerlines. A requirement on the minimum value of peaks, $\theta_v^{peak}$, and a requirement on the minimal allowed distance separating peaks, $\theta_d^{peak}$, are fulfilled during the sampling. The set of the must-pass nodes is denoted by $V^{mp}$, where $V^{mp} \subset V$ and $|V^{mp}|\ll|V|$.
Finding the optimal path that starts from $v_{st}$, and passes by all the must-pass nodes $V^{mp}$, and ends at $v_{ed}$ can be achieved by the constrained version of the Dijkstra’s algorithm\footnote{https://www.baeldung.com/cs/shortest-path-to-nodes-graph}. The complexity of the algorithm, $O(2^{|V^{mp}|}(|V|+|E| \cdot log(|V| \cdot 2^{|V^{mp}|})))$, increases exponentially with the number of the must-pass nodes, $|V^{mp}|$. Therefore, it is not feasible with a large number of must-pass nodes.

\paragraph{Graph Simplification}
Alternatively, we make a simplified graph $G'=(V',E')$, which consists of only the start, end, and the must-pass nodes, and then find a travelling salesman problem (TSP) solution on it. $V'=\{v'_m\}_{m=1}^{|V^{mp}|+2}=\{v_{st},v_{ed}\} \cup V^{mp}$ is a new set of nodes for the graph and $E'$ is a new set of edges.
An edge is constructed for every pair of nodes in G' and the edge cost $C_{m,n}'$ between nodes $v'_m$ and $v'_n$ is defined as:

\begin{equation}
    \label{eq:edge_cost_tsp}
    C_{m,n}'= 
    \begin{cases}
        \frac{cost(\Gamma_G^*(v'_m,v'_n))}{M}, & \text{if } d(v'_m,v'_n) \leq \delta\\
        \frac{d(v'_m,v'_n)}{\delta}, & \text{otherwise},
    \end{cases}
\end{equation}
where $M=\max_{\substack{(p,q) \in E' \\ d(v'_p,v'_q) \leq \delta}}  cost(\Gamma_G^*(v'_p,v'_q))$. $d(v'_m,v'_n)$ denotes the Euclidean distance between $v'_m$ and $v'_n$. $cost(\Gamma_G^*(v'_m,v'_n))$ denotes the cost of the shortest path between $v'_m$ and $v'_n$ on the initial graph $G$. Note that the shortest path is computed only for pairs within a distance of $\delta$, based on an assumption that a pair of nodes that are far from each other by the Euclidean distance would not be direct neighbors in a computed path. Otherwise, the Euclidean distance itself is used for calculating the edge cost. 

\paragraph{Solving a TSP}
We need a TSP solution that starts and ends at different given nodes. It can be achieved by adding a dummy node that has zero-cost edges with the start and end nodes, and infinity-cost edges with the remaining nodes. We use the nearest fragment operator~\cite{ray07} to solve the TSP.

\subsection{Evaluation Details}

The desired volume of each supervoxel and the compactness were set as $216mm^3$ and $0.01$ for Adaptive-SLIC~\cite{achanta12}. Thus, the number of the supervoxels, $N$, is decided depending on the size of an input volume. We adoped the same hyperparameters suggested in Ref.~\citenum{shin20} to train the network for small bowel segmentation. We used $\theta_v^{peak}=3mm$ and $\theta_d^{peak}=6mm$ for must-pass nodes sampling. $\delta=50mm$ was used for Equation~(\ref{eq:edge_cost_tsp}).

The predicted path is evaluated based on its overlap with the GT path. True positive (TP), false positive (FP), and false negative (FN) points are defined as in Ref.~\citenum{harten21}, with a static distance tolerance of $10 mm$. Precision and recall values are then computed from the TP/FP/FN values. We also compute the curve-to-curve distance, which is suggested in Ref.~\citenum{zhang18}. Our GT path is all connected from start to end of the small bowel compared to a set of segments of on average $88mm$ used in Ref.~\citenum{harten21}. To better understand the tracking capability of the proposed method, we provide the maximum length of the GT segment that is tracked without an error.

\section{RESULTS}\label{sec:results}
\subsection{Quantitative Evaluation}

Table~\ref{tab:quan_res} provides quantitative results of the baseline and the proposed methods. Although our graph is constructed based on the small bowel wall detection to prevent the tracking from penetrating a wall, simply computing the shortest path between start and end nodes, `Shortest Path', totally fails in terms of the recall. On the other hand, the proposed method has a much higher recall with the help of the must-pass nodes. It shows clear improvements for all metrics except for the precision. We note that, for the proposed method, the maximum length of the GT path that is tracked without an error per scan is above $800mm$ on average.

As mentioned in Section~\ref{subsec:graph_const}, small bowel segmentation is performed in the proposed method not only to remove the graph nodes that are located outside the small bowel but also to sample must-pass nodes within it. Therefore, the quality of the predicted segmentation affects the path tracking. The result based on the GT segmentation in place of the predicted segmentation is presented in Table~\ref{tab:quan_res}, implying that a better tracking can be achieved by improving the segmentation.

\begin{table}[t]
    \caption{Quantitative results for the baseline and the proposed methods. `Shortest Path' denotes computing the shortest path without the use of the must-pass nodes. The result of the proposed method, where the ground-truth segmentation is used in place of the predicted segmentation, is also presented. Refer to the text for the explanation on the evaluation metrics.} 
    \label{tab:quan_res}
    \begin{center}
    \begin{tabular}{c|c|c|c|c}
    Method & Precision (\%) & Recall (\%) & Curve-to-curve distance (mm) & Max. len. w/o error (mm) \\
    \hline
    Shortest Path & 97.8 & 9.6 & 28.0 $\pm$ 7.5 & 756.2 $\pm$ 341.5 \\
    \hline
    Ours & 92.5 & 92.7 & 4.2 $\pm$ 1.2 & 810.0 $\pm $193.6 \\
    \hline
    Ours w/ GT segm. & 98.1 & 96.0 & 3.3 $\pm$ 0.3 & 840.6 $\pm$ 273.1
    \end{tabular}
    \end{center}
\end{table}

\subsection{Qualitative Evaluation}

\begin{figure}[ht]
	\centering
	\begin{minipage}{1\textwidth}
        \subfigimg[width=0.33\textwidth,pos=ll, font=\color{black}]{A}{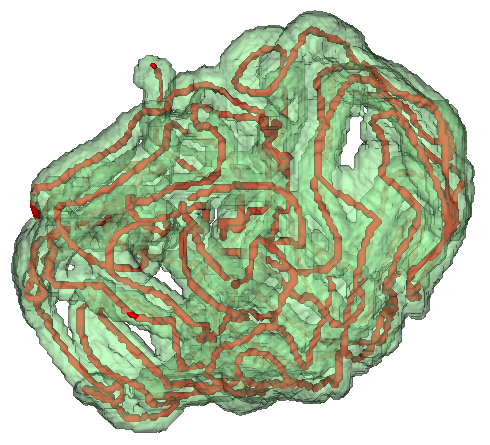}
        \hspace{-0.25cm}
        \subfigimg[width=0.33\textwidth,pos=ll, font=\color{black}]{B}{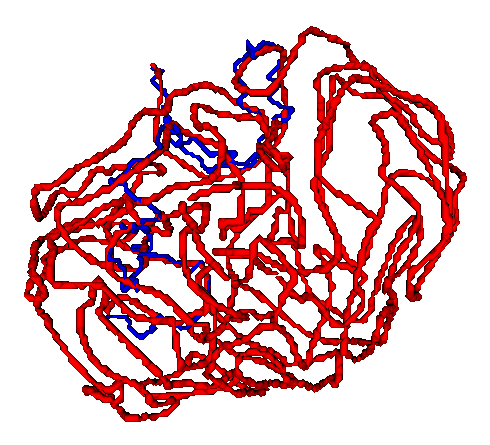}
        \hspace{-0.25cm}
        \subfigimg[width=0.33\textwidth,pos=ll, font=\color{black}]{C}{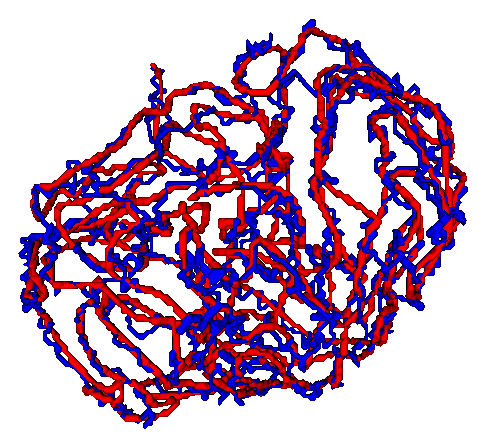}
    \end{minipage}
    \begin{minipage}{1\textwidth}
        \subfigimg[width=0.33\textwidth,pos=ll, font=\color{white}]{D}{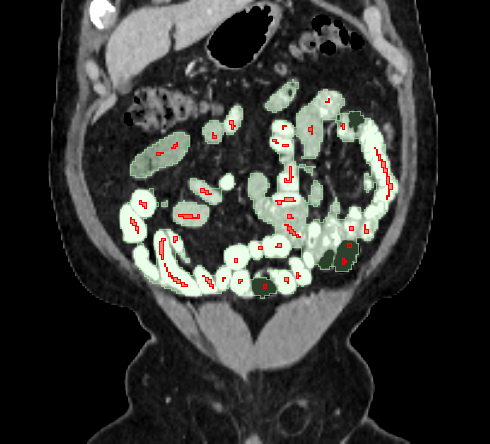}
        \hspace{-0.25cm}
        \subfigimg[width=0.33\textwidth,pos=ll, font=\color{white}]{E}{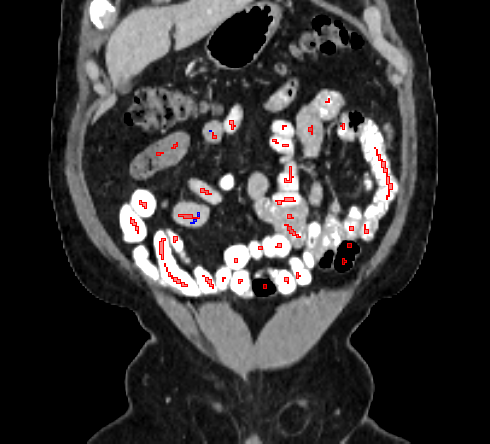}
        \hspace{-0.25cm}
        \subfigimg[width=0.33\textwidth,pos=ll, font=\color{white}]{F}{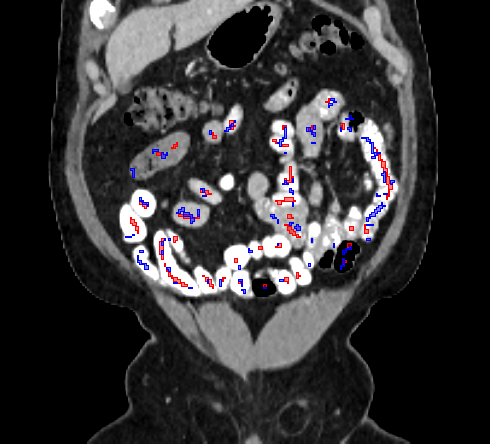}
    \end{minipage}
	\caption{Example path tracking results. (A) Ground-truth (GT) path annotation (red) overlayed on the predicted small bowel segmentation (green). (B) Result corresponding to `Shortest Path' in Table~\ref{tab:quan_res}. (C) Result of the proposed method. In (B) and (C), each result (blue) is compared with the GT (red). (D-F) Coronal view results corresponding to (A-C).}
	\label{fig:qual_res}
\end{figure}

Fig.~\ref{fig:qual_res} shows example path tracking results. Observing the CT scans, the contact between different parts of the small bowel, indistinct appearance of the wall, and the heterogeneity of the lumen make the problem difficult. The trivial solution generated by the baseline method proves it. The proposed method better covers the entire course of the small bowel by the help of the must-pass nodes.

\section{CONCLUSION}\label{sec:conclusion}

We have presented a new graph-based method for small bowel path tracking. We find the minimum cost path on a graph constructed based on the wall detection. An explicit constraint on the resulting path, which is implemented by the must-pass nodes, is also applied to better cover the entire course of the small bowel. The tracked path provides a better understanding on the small bowel structure than the segmentation, enabling precise localization of diseases along the course of the small bowel. Especially, it could help interpreting a CT scan before using capsule endoscopy, and preoperative planning by better visualization. In future work, we plan to include CT scans with different contrast media to increase the applicability of the proposed method.

\acknowledgments 
 
We thank Dr. James Gulley for patient referral and for providing access to CT scans. This research was supported by the Intramural Research Program of the National Institutes of Health, Clinical Center.


\end{document}